\documentstyle[preprint,aps,epsfig]{revtex}   
 
\begin{document}   

\tightenlines

\title{Transport in random quantum dot superlattices}  

\author{I.\ G\'{o}mez, F.\ Dom\'{\i}nguez-Adame and E.\ Diez}

\address{GISC, Departamento de F\'{\i}sica de Materiales, Universidad
Complutense, E-28040 Madrid, Spain}

\author{P.\ Orellana}

\address{Departamento de F\'{\i}sica, Universidad Cat\'{o}lica del Norte, 
Casilla 1280, Antofagasta, Chile}

\date{\today}   
   
\maketitle

\begin{abstract} 

We present a novel model to calculate single-electron states in random quantum
dot superlattices made of wide-gap semiconductors. The source of disorder comes
from the random arrangement of the quantum dots (configurational disorder) as
well as spatial inhomogeneities of their shape (morphological disorder). Both
types of disorder break translational symmetry and prevent the
formation of minibands, as occurs in regimented arrays of quantum dots. The
model correctly describes channel mixing and broadening of allowed energy bands
due to elastic scattering by disorder.  

\end{abstract}   
   

\narrowtext

\section{Introduction}

Latest advances in nanotechnology make it possible to growth quantum dot~(QD)
superlattices~\cite{Liu99a,Liu99b}. In view of the analogy between atoms and
QDs, it is expected that strongly confined levels overlap when QDs are closely
packed. Although this analogy cannot be complete since carriers in QDs are
influenced by phonons, defects and interface states, the resulting collective
levels will depend on the particular arrangement of QDs. In this sense, QD
arrays grown by molecular beam epitaxy can be completely
random~\cite{Tanaka99,Sharma99}, partially regimented~\cite{Liu99a,Liu99b} or
may be regularly stacked (high regimentation)~\cite{Springholz98}. Electronic
states in highly regimented QDs are adequately described within
envelope-function approximation with a three dimensional Kronig-Penney model
and the occurrence of miniband have been established~\cite{Lazarenkova01}.
However, the lack of periodicity observed in random QDs superlattices demands
different approaches.

In the present work we introduce a two-dimensional effective-mass model to
study the effects of scattering by disorder on electron transmission through
random quantum-dots superlattices. To this end, the Ben Daniel-Duke equation is
discretized, boundary conditions are discussed and scattering solutions are
found by means of the transfer-matrix method for {\em any\/} arbitrary array of
QDs. The model is worked out in a two-dimensional space for computational
limitations, although it will be clear that generalization to three dimensions
is rather straightforward. Finally, we present the numerical results for the 
transmission coefficient through random arrays of coupled QDs and the main
conclusions of the work.

\section{Model}  

We consider the Ben Daniel-Duke equation with constant effective mass $m^{*}$
at the $\Gamma$ valley for the electron envelope function $\chi(y,z)$. In order
to find numerically the single electron states, the whole sample is divided
into three different regions, namely left (I) and right (III) contacts and the
random QDs (II), where scattering by disorder takes place.  Figure~1 shows a
schematic view of the three spatial regions. We then consider a mesh with
lattice spacings $a_y$ and  $a_z$ in the $y$ and $z$ directions, respectively.
Defining $t_y \equiv -\hbar^{2}/(2m^{*}a_y^2)$ and $t_z\equiv
-\hbar^{2}/(2m^{*}a_z^2)$, we obtain the following discrete effective-mass
equation
\begin{equation} 
\label{discrete}
t_z(\chi_{n+1,m}+\chi_{n-1,m})+t_y(\chi_{n,m+1}+\chi_{n,m-1}) 
+(U_{n,m}-2t_z-2t_y)\chi_{n,m} = E\chi_{n,m}.
\end{equation}
The potential term $U_{n,m}$ in Eq.~(\ref{discrete}) is given by the
conduction-band edge energy at the point $(na_y,ma_z)$. Therefore, disorder
enters the equation through this diagonal term. Contacts are characterized by
flat band conditions, $U(n,m)=0$, in the absence of applied electric field. 
Clearly, the effects of the applied field can be easily taken into account
within the present approach by adding a linearly varying potential of
the form $U_{\mathrm{F}}(n,m)=-eFma_z$, $F$ being the applied field.

\section{Transfer matrix formalism}  

In order to solve the tight binding-like equation~(\ref{discrete}) we use the
transfer matrix method based on the solutions calculated for each slide of the
system along the $z$ direction. For the sake of simplicity in the calculation,
we define $\vec{\phi}_{n}$ ($n=0,1,\ldots,N+1$) as a vector whose components
are  $\phi_{n}^{m}\equiv \chi_{n,m}$ (m=1,\ldots,M). Here $M$ and $N+1$ are the
number of mesh divisions in the $y$ and $z$ directions, respectively. Thus, 
Eq.~(\ref{discrete}) can be cast in a more compact form
\begin{equation} 
\label{promotion}
\left(
\begin{array}{c}
\vec{\phi}_{n-1} \\
\vec{\phi}_{n}
\end{array} 
\right) =
\left(
\begin{array}{cc}
t_{z}^{-1}(E {\cal I} - {\cal M}_n) & - {\cal I}\\
{\cal I} & {\cal O}
\end{array} 
\right)
\,
\left(
\begin{array}{c}
\vec{\phi}_{n} \\
\vec{\phi}_{n+1}
\end{array}
\right),
\end{equation}
where $\cal I$ and $\cal O$ are the $M\times M$ identity and null matrices
respectively. The matrix ${\cal M}_n$ splits into the form ${\cal M}_{n}  =
{\cal R}_{n} + {\cal B}_{n}$. The diagonal elements of the tridiagonal matrix
${\cal R}_{n}$ are $({\cal R}_{n})_{mm}= U_{n,m}-2t_z-2t_y$ while nonvanishing
off-diagonal elements equal $t_y$. The matrix ${\cal B}_{n}$ depends on the
boundary conditions to be specified later.

We can obtain the expression of the envelope function amplitudes in the left
contact as a function of the amplitudes in the right one
\begin{equation} 
\label{transfer}
\left(
\begin{array}{c}
\vec{\phi}_{0} \\
\vec{\phi}_{1}
\end{array} 
\right) =
{\cal T}^{(N)}
\,
\left(
\begin{array}{c}
\vec{\phi}_{N} \\
\vec{\phi}_{N+1}
\end{array}
\right),
\quad
{\cal T}^{(N)} \equiv \prod_{n=1}^{N}
\left(
\begin{array}{cc}
t_{z}^{-1}(E {\cal I} - {\cal M}_n) & - {\cal I} \\
{\cal I} & {\cal O}
\end{array} 
\right),
\end{equation}
where ${\cal T}^{(N)}$ is the transfer matrix for the heterostructure.

\section{Scattering solutions}  

The envelope functions within the contacts will be determined by the boundary
conditions. These boundary conditions are open in the $z$ direction, and
periodic on each slide, namely in the $y$ direction. Consequently, all elements
of ${\cal R}_{n}$ vanish except $({\cal R}_{n})_{1M}=({\cal R}_{n})_{M1}=t_y$.
The former conditions imply plane wave solutions in the $z$ axis, and the
latter yield an energy discretization on $y$. As a consequence, this
discretization results in a number of transverse channels equal to the number
of points in the transverse mesh direction. Considering $U_{n,m}$  to be
constant at the contacts, that is, considering perfect leads so no voltage drop 
occurs within regions I and III, and  setting $\chi_{n,1}=\chi_{n,M+1}$, a 
particular solution of Eq.~(\ref{discrete}) is given by
\begin{mathletters}
\begin{eqnarray} 
\label{solutionI}
\chi_{n,m}^{l}&=&\frac{1}{\sqrt{{\cal N}_l}}
\exp\left(i\,\frac{2\pi l}{M}m\right)
\exp\left(ik_l a_z n\right) \nonumber \\
&+& \sum_{j=1}^{M}\widehat{r}_{lj} \frac{1}{\sqrt{{
\cal N}_j}}\exp\left(i\,\frac{2\pi j}{M}m\right)
\exp\left(-ik_ja_z n\right), \quad (m,n) \in \mathrm{I}.
\end{eqnarray}
at the left contact while at the right contact we have
\begin{equation} \label{solutionIII}
\chi_{n,m}^{l}=
\sum_{j=1}^{M} \widehat{t}_{lj} \frac{1}{\sqrt{{\cal N}_j}}\exp\left(i\,
\frac{2\pi j}{M}m\right) \exp\left(i k_j a_z n\right),
\quad (m,n) \in \mathrm{III}.
\end{equation}
\end{mathletters}
where the normalization constant is choosen to ensure that all the {\it
propagating} modes carry the same current
$$
{\cal N}_j = \frac{1}{a^2_y}\,\sin^2 \left( \frac{2 \pi j}{M} \right)
+\frac{1}{a^2_z}\,\sin^2 \left( k_j a_z \right)
$$
and 
$$
k_j = \frac{1}{a_z} \cos^{-1} \left\{ \frac{1}{2t_z} \left[ E - 2t_y 
\left( \cos \frac{2\pi j}{M} -1 \right) \right] + 1 \right\}.
$$
These expressions remain valid for an applied electric field $F$ provided the
electronic momentum $k_j$ is substituted within region $III$ in the following 
way
\begin{eqnarray}
k_j & \leftrightarrow & q_j \nonumber \\
q_j & = & \frac{1}{a_z} \cos^{-1} \left\{ \frac{1}{2t_z} \left[ E + eV - 2t_y 
\left( \cos \frac{2\pi j}{M} -1 \right) \right] + 1 \right\}.
\end{eqnarray}
Here $V=FL$, where $L$ is the lenght of region $II$.

The matrices $\widehat{r}$ and $\widehat{t}$ appearing in equations
(\ref{solutionI}) and (\ref{solutionIII})
are the {\em reflection\/} and {\em transmission\/} matrices and they are
responsible for the channel {\em mixing\/} due to scattering events. Thus,
$\widehat{r}_{ij}$ represents the probability amplitude for an electron
impinging in channel $i$ to be reflected into the chanel $j$. Note that the
solution  within region II is unknown. Actually, we are not interested in this
solution since all the physics of the scattering problem is contained in the
{\em mixing\/} matrices $\widehat{t}$ and $\widehat{r}$. In particular, we can
compute the conductance. From the Landauer-B\"{u}ttiker
formalism\cite{Landauer57}, the zero temperature two-leads multichannel
conductance can be calculated using de Fisher-Lee formula~\cite{Fisher81}
\begin{equation} 
\label{conductance}
G = \frac{2e^2}{h} \,\text{Tr} (\widehat{t}^{\dag}\widehat{t}),
\end{equation}
where $\text{Tr}$ stands for the trace of the matrix.

\section{Configurational and morphological disorder}  

In order to describe epitaxially random QD superlattices, we consider they are
arranged on a nonregular lattice (configurational disorder). But, in addition,
there exists another source of disorder due to spatial inhomogeneities that
make the shape to be slightly different (morphological disorder). To mimic both
types of disorder we assume an array of rectangular QDs, randomly displaced
from the regular lattice sites, whose size also change randomly from dot to dot
(see Fig.~\ref{fig1}). To avoid the profusion of free parameters, we consider
that the energy of the confinement potential $\Delta E_c$ provided by the high
bandgap semiconductor is the same for every QD. This is not a serious
shortcoming since spatial inhomogeneities of the conduction-band offset and
fluctuations of individual QDs shapes yield essentially the same results. In
addition, our model could easily deal with nonconstant values of $\Delta E_c$ if
further improvements are required.

We will separate the effects of configurational and morphological disorder for
the sake of clarity. Configurational disorder means that every QD shifts its
position $\delta{\mathbf r}=(\delta y,\delta z)$ while its size $d_y\times d_z$
remains unchanged (see Fig.~\ref{fig2}). Here $\delta y$ and $\delta z$ are
random uncorrelated variables with zero mean and distributed according to box
probability functions of width $W_y$ and $W_z$, respectively. To simulate the
change of the shape (morphological disorder) we consider that the QD is
enlarged along the $Z$ axis an amount $\delta \zeta$ while its center stays on
a regular lattice (see Fig.~\ref{fig2}), where $\delta \zeta$ is uniformly
distributed with zero mean and width $W_\zeta$.

\section{Results}  

As a working example, we have performed several numerical calculations in order
to study the effect of both configurational and morphological disorder over the
conductance of random QDs superlattices made of GaAs-In$_x$Ga$_{1-x}$As
heterojunctions. We have considered regimented and disordered $4\times 4$ arrays
of QDs to elucidate the effects of randomness. The conduction-band offset,
$\Delta E_c$, is taken to be $70\%$ of the difference of the gaps $\Delta E_g$
in strained GaAs-In$_x$Ga$_{1-x}$As heterojunctions, where $\Delta E_g =
1.45x\,$eV. For definiteness we set $x=0.35$ and consequently $\Delta E_c =
0.35\,$eV. In addition, since we are mainly interested in the effects of the
coupling between the QDs through the high bandgap semiconductor rather in the
confined levels of individual QD, we have taken $m^{*}=0.067$ in units of the
bare electron mass, corresponding to the embedding semiconductor. Let us
mention that the the model can be easily generalized to include a different
effective mass inside the QDs. The size of the regimented QDs is $d_y\times
d_z$ with $d_y=8.0\,$nm and $d_z=1.6\,$nm. The separation between centers in
the regimented array is $14.0$ and $6.8\,$nm along the $Y$ and $Z$ axes,
respectively. The number of mesh points along the two spatial directions are
$M=50$ and $N+1=39$. 

As typical results of our simulations, Fig.~\ref{fig3} shows the conductance
versus Fermi energy, measured from the conduction-band edge in
In$_x$Ga$_{1-x}$As, in the absence of applied field. Solid line corresponds to
regimented QDs. The coupling between the QDs splits the energy levels and
results in the formation of minibands~\cite{Lazarenkova01}. From
Fig.~\ref{fig3} we observe the occurrence of two well-defined minibands below
the barrier when disorder is absent, in agreement with previous
results~\cite{Lazarenkova01}. Each band is characterized by four main
conductance peaks and each peak is the convolution of four peaks that cannot be
resolved except for the lower one in the higher miniband (see inset of
Fig.~\ref{fig3}). 

Transport though the miniband changes as soon as some degree of randomness is
considered in the model, as expected. Mean size and separation between QDs as
well as fluctuations around those values strongly depend on the growth
conditions (e.g. growth temperature) and subsequent thermal
treatments~\cite{Sharma99}. As an example, Fig.~\ref{fig3} shows the
conductance for configurational ($W_y=2.0\,$nm and $W_z=1.2\,$nm) and
morphological ($W_\zeta=0.8\,$nm) disorder, obtained by averaging over $100$
realizations of the disorder. As a main point, we notice that electronic states
in the array of random QDs behaves like an amorphous material since the
conductance strongly decreases while the allowed energies broadens due to the
fluctuations of QD energy levels for each realization of the disorder.

Concerning the effects of a uniform applied electric field, we have computed
the conductance for a given Fermi energy ($E=0.27\,$eV) as a
function of the applied bias $V$. The potential was assumed to drop uniformly
across region~II. Figure~\ref{fig4} shows the conductance versus applied bias
for regimented as well as random QD superlattices. Regimented QD superlattices
present three well-defined negative $dG/dV$ regions due to
resonant tunneling through the QDs. The observed peak-to-valley ratios become
worse as soon as disorder is included in the calculation, especially in the
case of configurational disorder.

\section{Conclusions}

In summary, we have presented a method to study electron transport through
random QD superlattices. The method is based on the transfer matrix formalism
applied to the discrete Ben Daniel-Duke Hamiltonian for the electron envelope
function. A careful analysis of the scattering solutions under appropriate
boundary conditions (periodic and open along the lateral and longitudinal
directions, respectively) allows us to obtain the two-channel conductance. For
regimented QDs (regular array of QDs) the conductance shows clear signatures of
the miniband structure, as previously predicted by a Kronig-Penney model for
strongly coupled QDs~\cite{Lazarenkova01}. However, the novelty of the model
lies in the fact that random QD superlattices can also be studied. To this end,
two models of disorder (configurational and morphological) have been 
introduced. Disorder reduces the conductance due to Anderson localization of the
envelope functions and broadens the allowed energy bands. The characteristic
$G$--$V$ presents several regions of negative $dG/dV$, although
the peak-to-valley ratios strongly decrease due to disorder.

\acknowledgments

Work in Madrid was supported by DGI-MCyT (Project~MAT2000-0734) and CAM
(Project~07N/0075/2001). P.\ Orellana would like to thank Milenio ICM P99-135-F
and C\'{a}tedra Presidencial de Ciencias for financial support.

\begin{figure}
\caption{Schematic view of the sample. Regions I and III are the electrical
leads of the sample (contacts) and electrons undergo scattering processes only 
at region II.}
\label{fig1}   
\end{figure}

\begin{figure}
\caption{Schematic view of configurational and morphological disorder. In the
former case the QD shift its position an amount $\delta{\mathbf r}$ while
its nominal size $d_y \times d_z$ remains unchanged. In the later case, 
the center of the QD does not change while its size along the $Z$ axis 
increases or decreases an amount $\delta \zeta$.}
\label{fig2}   
\end{figure}

\begin{figure}
\caption{Conductance versus energy for a 2D ordered array of $4\times 4$ QDs 
made of In$_x$Ga$_{1-x}$As in GaAs (solid line) with no applied electric field.
The inset shows an enlarged view of the lower conductance peak of the second
miniband. Results are compared to random $4\times 4$ arrays with
configurational (dotted line) and morphological (dashed line) disorder.}
\label{fig3}   
\end{figure}

\begin{figure}
\caption{Conductance versus applied bias for a 2D ordered array of $4\times 4$
QDs  made of In$_x$Ga$_{1-x}$As in GaAs (solid line), when the incoming 
electron energy is $E=0.27\,$eV. Results are compared to random $4\times 4$
arrays with configurational (dotted line) and morphological (dashed line)
disorder.}
\label{fig4}   
\end{figure}

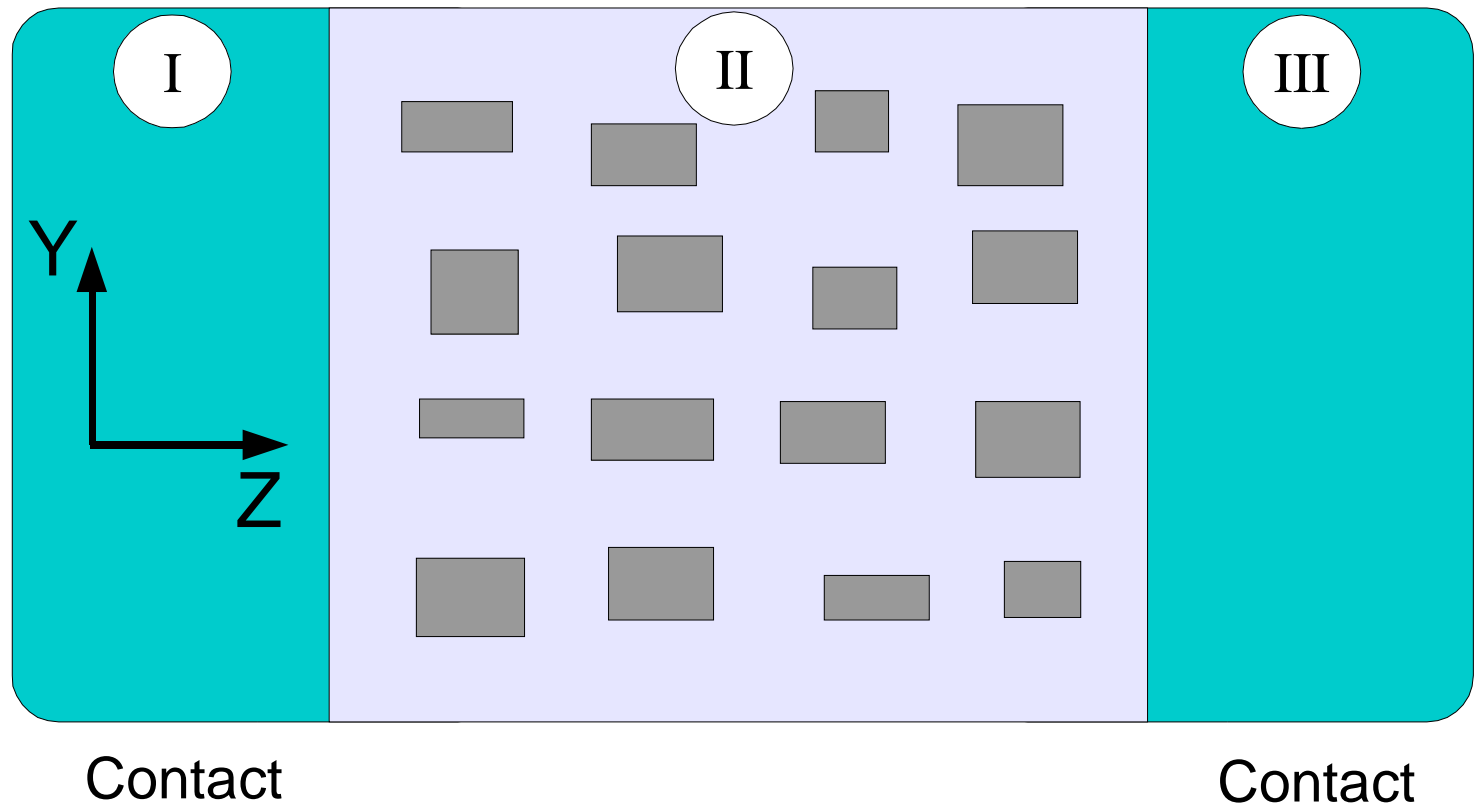
\end{document}